\documentclass[5p]{elsarticle}

\usepackage{amssymb}
\usepackage{amsmath,mathrsfs}
\usepackage{color,soul}

\usepackage{extarrows}

\usepackage{verbatim}

\usepackage{lineno}

\usepackage{siunitx}

\usepackage{stfloats}

\usepackage{bm}

\usepackage{graphicx}

\usepackage{caption}
\usepackage{subcaption}

\usepackage{multicol}

\usepackage{multirow}

\usepackage{tabularx}

\usepackage[version=4]{mhchem}

\usepackage[toc,page]{appendix}

\usepackage{bm}

\usepackage{float}

\usepackage{empheq}

\usepackage{threeparttable}

\usepackage{sectsty}
\usepackage[toc,page]{appendix}
\usepackage[colorlinks,linkcolor=blue,anchorcolor=green,citecolor=blue]{hyperref}

\newcommand{\figref}[1]{Fig.~\ref{#1}}
\newcommand{\tabref}[1]{Table.~\ref{#1}}
\renewcommand{\eqref}[1]{Eq.~$($\ref{#1}$)$}

\journal{Journal of \LaTeX\ Templates}

\begin{document}
\newcommand*{\E}[1]{\mathop{}\!\times 10^{#1}}
\newcommand*{\fed}[1]{f_\mathrm{#1}}
\newcommand*{\expect}[1]{\langle{#1}\rangle}

\def\diff{\mathrm{d}}

\newcommand*{\varid}[2]{\mathop{}\!\frac{\delta #1}{\delta #2}}

\def\mag{\mathbf{m}}
\def\mv{\mathbf{m}}
\def\eau{\mathbf{u}}
\def\pos{\mathbf{r}}
\def\hh{\mathbf{h}}
\def\H{\mathbf{H}}
\def\Hv{\mathbf{H}}
\def\M{\mathbf{M}}
\def\J{\mathbf{J}}
\def\B{\mathbf{B}}

\def\H{\mathbf{H}}

\def\Hn{H_\mathrm{n}}
\def\Hp{H_\mathrm{p}}
\def\Hc{H_\mathrm{c}}

\def\Hsw{H_\mathrm{sw}}
\def\Hext{\mathbf{H}_\mathrm{ext}}
\def\hext{{H}_\mathrm{ext}}
\def\hc{{H}_\mathrm{c}}

\def\Msat{M_\mathrm{s}}
\def\Hdm{\mathbf{H}_\mathrm{dm}}

\def\Ae{A_\mathrm{ex}}
\def\Ku{K_{\mathrm{u}}}
\def\sigdw{\sigma_\mathrm{dw}}
\def\lex{l_\mathrm{ex}}

\def\dfedfdm{\frac{\delta \mathcal{F}}{\delta \mathbf{m}}}

\def\ngl{l_\mathrm{nc}}
\def\fv{\varphi_\mathrm{nc}}

\begin{frontmatter}

\title{Influence of amorphous phase on coercivity in \ce{SmCo_{5}}-Cu nanocomposites}


\author[mymainaddress]{Franziska Staab}\corref{mycorrespondingauthor}

\author[mysecondaryaddress]{and Yangyiwei Yang\corref{mycorrespondingauthor}}
\cortext[mycorrespondingauthor]{Authors contributed equally; Corresponding author}
\ead{yangyiwei.yang@mfm.tu-darmstadt.de}
\author[mysecondaryaddress]{Eren Foya}
\author[mymainaddress]{Enrico Bruder}
\author[mythirdaddress,myfourthaddress]{Benjamin Zingsem}
\author[myfifthaddress]{Esmaeil Adabifiroozjaei}
\author[mysixthaddress]{Konstantin Skokov}
\author[mythirdaddress]{Michael Farle}
\author[myfourthaddress]{Rafal E. Dunin-Borkowski}
\author[myfifthaddress]{Leopoldo Molina-Luna}
\author[mysixthaddress]{Oliver Gutfleisch}
\author[mysecondaryaddress]{Bai-Xiang Xu}
\author[mymainaddress]{Karsten Durst}

\address[mymainaddress]{Physical Metallurgy (PhM), Materials Science Department, Technical University of Darmstadt, Alarich-Weiss-Str. 2, 64287 Darmstadt, Germany}
\address[mysecondaryaddress]{Mechanics of Functional Materials (MFM), Materials Science Department, Technical University of Darmstadt, Otto-Berndt-Str. 3, 64287 Darmstadt, Germany}
\address[mythirdaddress]{Faculty of Physics and Center for Nanointegration (CENIDE), University of Duisburg-Essen, 47057 Duisburg, Germany}
\address[myfourthaddress]{Ernst Ruska-Centre for Microscopy and Spectroscopy with Electrons and Peter Grünberg Institute, Forschungszentrum Jülich GmbH, 52425 Jülich, Germany}
\address[myfifthaddress]{Advanced Electron Microscopy (AEM), Materials Science Department, Technical University of Darmstadt, Alarich-Weiss-Str. 2, 64287 Darmstadt, Germany}
\address[mysixthaddress]{Functional Materials (FM), Materials Science Department, Technical University of Darmstadt, Alarich-Weiss-Str. 16, 64287 Darmstadt, Germany}

\begin{abstract}
Severe plastic deformation of powder blends consisting of \ce{SmCo_{5}}-Cu results in magnetically hardened nanocomposite bulk materials. The microstructure is continuously refined with increasing torsional deformation, yet, coercivity saturates at a certain level of strain. Transmission electron microscopy (TEM) investigation of the microstructure reveals a partial amorphization of the \ce{SmCo_{5}} phase due to high-pressure torsion by 20 applied rotations. In this amorphous matrix nanocrystals are embedded. The effect of these experimentally observed microstructural features on the magnetic properties are investigated by micromagnetic simulations, which show that an increasing volume fraction of nanocrystals is beneficial for higher coercivities. For a fixed volume fraction of nanocrystals the simulations reveal an increasing coercivity with decreasing the size of the nanocrystals due to increasing number of interfaces acting as pinning sites. 
Furthermore, our micromagnetic simulations disclose the  mechanisms of the saturation and decline of magnetic hardening due to the strain induced by high-pressure torsion. The calculated coercivity fits very well to the experimentally observed coercivity of $\hc=1.34 \si{T}$. The knowledge can also be used to develop and provide optimization strategies from the microstructure perspective.
\end{abstract}


\end{frontmatter}

As the demand for renewable energy sources increases, the demand for high-performance permanent magnets is also rising \cite{GUT11}. In order to fulfill the requirements for suitable hard magnetic materials, good intrinsic magnetic properties need to be combined with optimized extrinsic magnetic properties, notably the remanent magnetization $M_\mathrm{r}$ and the coercivity $\Hc$. Therefore, the micro- or nanostructure need to be adjusted accordingly. The conventional processing techniques for permanent magnets are based on a powder-metallurgical sintering route \cite{SKO18}. It has been shown that the coercivity of such magnets usually does not exceed $20-30\,\%$ of the theoretical prediction of $\Hc = {2\Ku}/{\mu_0 \Msat}$ ($\Ku$ as the 1$^\mathrm{st}$ magnetocrystalline uniaxial anisotropy constant, $\mu_0$ as the vacuum permeability, and $\Msat$ as the saturated magnetization) given by the Stoner-Wohlfarth model for a magnetic field applied along the easy direction \cite{STO48,SKO18}. Hence, it is challenging to develop a technology to adjust the nano- or microstructure of bulk magnetic materials.
\\ \\
Severe plastic deformation by high pressure torsion (HPT) being applied to powder blends has recently been shown as a promising approach for the production of textured nanocomposite hard magnetic materials \cite{STA23, WEI20, WEI22, WEI23}. Applying the HPT process to powder blends allows the free selection of the magnetic phase and the grain boundary phase, {without limitation of phase formations as in the case of the powder-metallurgical sintering route.} In addition, HPT enables the adjustment of the microstructure by parameter variation such as the applied strain, pressure and temperature \cite{STA23, WEI20, WEI23}. However, examination the microstructural effects on the magnetization reversal behavior on small length scales, is very difficult to access experimentally.

Recently, micromagnetic modeling has been very successful in simulating magnetization dynamics under various stimuli such as magnetic, elastic, and thermal fields. It has been widely employed in examining the role of microstructural features on magnetic reversal mechanisms, that is domain nucleation and domain wall pinning \cite{yi2016micromagnetic, duerrschnabel2017atomic}. Up to date, micromagnetic simulations have been successfully performed on various microstructures of permanent magnets, notably, the nanoscopic granular \cite{kim2019effect, bautin2017magnetic, HONO20186} and cellular structures \cite{duerrschnabel2017atomic}, in which various structural features, such as the grain shape \cite{yi2016micromagnetic, kovacs2020computational, fuentes2017micromagnetic}, aspect ratio \cite{tang2020relationship}, and misalignment \cite{kim2019effect} have been examined and discussed. However, the microstructural evolution of nanocomposites, produced by severe plastic deformation via HPT, has been rarely modeled. Here we will show that with the help of our micromagnetic simulations using the experimentally observed microstructure, we can explain the increase and subsequent decrease of the coercivity in HPT processed \ce{SmCo_{5}}-Cu powder blends.

In this work, the nanocomposite of \ce{SmCo_{5}} and Cu was produced by high-pressure torsion of a powder blend consisting of 80~wt.\% intermetallic fine-grained powder of \ce{SmCo_{5}} (Alfa Aesar) and 20~wt.\% of Cu powder (99.9~\%, Alfa Aesar). First the powder was precompacted by HPT by the aid of a Cu-ring, placed on the lower anvil. For this the powder was filled into the Cu-ring and compacted at room temperature at a nominal pressure of 4.5~\si{GPa} by ten oscillations of \textpm 5\textdegree\ each. In a second HPT step the consolidated discs with a diameter of 10~\si{mm} and an initial height of 1.4~\si{mm} were deformed by 1, 20 and 100 rotations, respectively, with a rate of 2~rpm at a nominal pressure of 7.6~\si{GPa}. 

\begin{figure*}[h]
 \centering
 \includegraphics[width=0.85\textwidth]{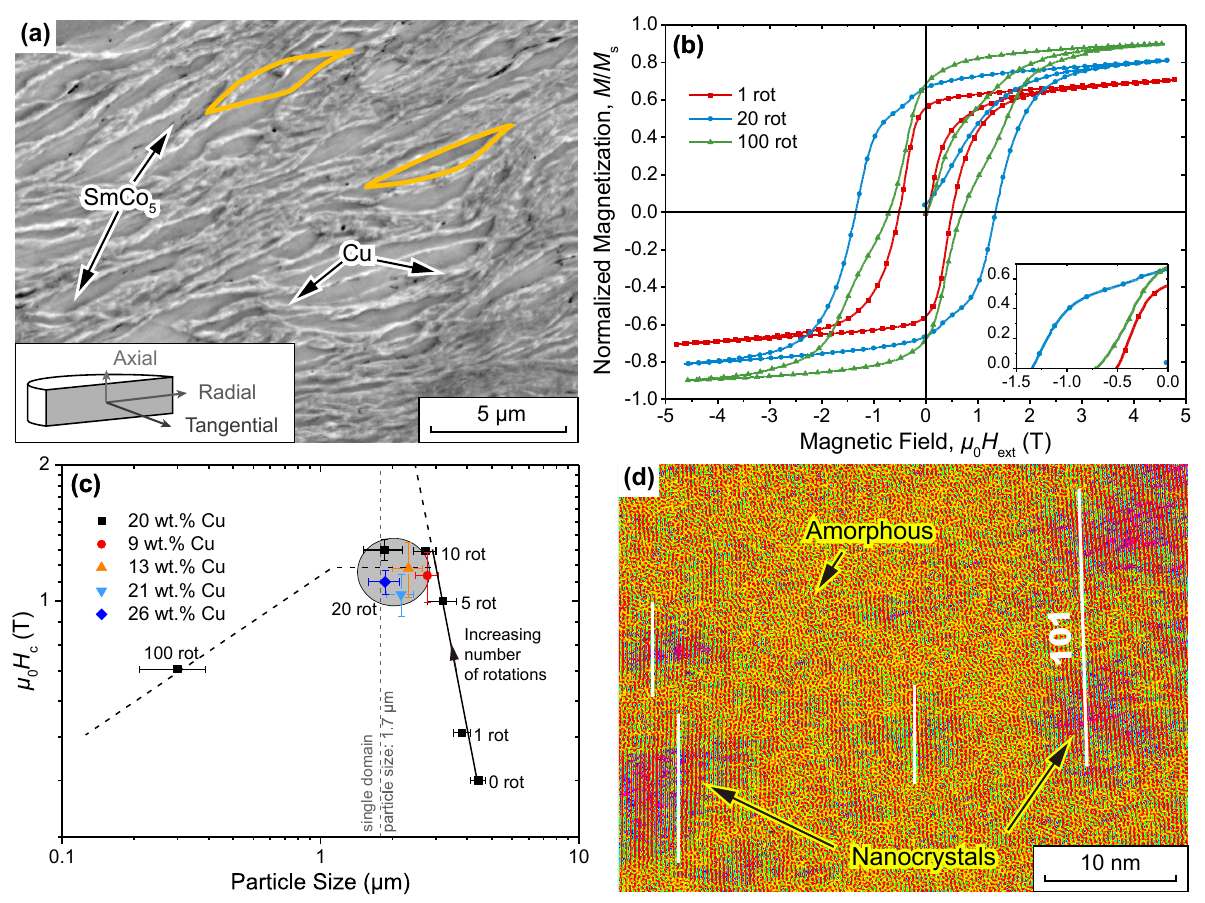}
 \caption{(a) SEM image of the cross section of a HPT deformed sample subjected to 20 rotations with marked elongated \ce{SmCo_{5}} particles, adapted from \cite{STA23} with permission, (b) Room temperature magnetic hysteresis curves of the samples after 1, 20 and 100 rotations measured in axial direction. Here the saturation magnetization $\Msat=648.6~\si{kA/m}$, (c) coercivity as a function of the particle size, (d) IFFT image with marked nanocrystals along the [101] direction, adapted from \cite{STA23} with permission.}
\label{Fig_Microstructure_Experiment}
\end{figure*}

For microstructural analyses the cross section of the discs were investigated using a high-resolution scanning electron microscope (TESCAN Mira 3) by backscattered electrons (BSE). A line interception method was applied for quantitative evaluation of the \ce{SmCo_{5}}-particle size following the same approach as described in \cite{STA23}. In a region showing strong refinement of the microstructure, TEM analyses were performed on the sample subjected to 20 rotations on a JEM 2100F TEM operated at 200 \si{kV}. The magnetic hysteresis curve $M(H)$ was measured in a magnetic field up to \textpm 5 \si{T} applied along the axial direction (height of the disc) by a Physical Property Measurement System (PPMS, Quantum Design) with a Vibrating Sample Magnetometer (VSM) at 300\si{K} .
\\ \\
\figref{Fig_Microstructure_Experiment}a shows the microstructure of the \ce{SmCo_{5}}-Cu nano-composite subjected to 20 rotations obtained by backscatterd electron (BSE) imaging. The \ce{SmCo_{5}} particles show a strong elongation perpendicular to the direction of the applied pressure and the particles are surrounded by the diamagnetic fcc Cu-phase. In \figref{Fig_Microstructure_Experiment}b the demagnetization corrected hysteresis curves measured in axial direction of the samples subjected to 1, 20 and 100 rotations are depicted. For the calculation of the internal magnetic field a density of 8.584~g/cm\textsuperscript{3} for 20~wt.\% of Cu was assumed. Comparing the samples subjected to 1 and 20 rotations a strong increase of the coercivity and of the saturation magnetization can be seen. The saturation magnetization continues to increase for the sample subjected up to 100 rotations, while the coercivity decreases from 1.34 \si{T} after 20 rotations to 0.51 \si{T}.

In \figref{Fig_Microstructure_Experiment}c the coercivity is plotted as a function of the grain size. In addition to $H_c$ of the samples subjected to 1, 20 and 100 rotations, the diagram also contains $H_c$ of other samples prepared in \cite{STA23}. $H_c$ increases with a reduction in particle size for the samples subjected to up to 10 rotations. This increase is related predominantly to the particle refinement in combination with thin Cu-layers in between the ferromagnetic particles with a sufficient thickness for magnetic decoupling \cite{STA23, NEY99}. With further increase of strain $H_c$ remains nearly constant even though the particle size further decreases and reaches the single domain size of $l=1.7~\si{\micro m}$ \cite{GUT00} for the sample subjected to 20 rotations. Therefore, a further increase of the coercivity is expected, indicated by the dashed line. After 100 rotations the particle size is further diminished and the coercivity strongly decreases as already seen in \figref{Fig_Microstructure_Experiment}b.

TEM investigations by inverse fast Fourier transform analyses (\figref{Fig_Microstructure_Experiment}d) reveal a partial amorphization of the \ce{SmCo_5} phase, whereas nanocrystals of 10 to 50 nm in size are embedded in an amorphous matrix. The volume fraction of the nanocrystals was determined to $\fv=0.26$ for the sample subjected to 20 rotations. In addition to the selective TEM observation, XRD data, shown for as-HPT deformed samples up to 20 rotations in \cite{STA23}, indicate a partial amorphization on a macroscopic length scale by a clear broadening of the reflexes with increasing number of rotations. The amorphous phase is believed to play a crucial role in the saturation behavior of coercivity and possibly also in the further decrease in magnetic hardening with straining. 

\begin{figure*}[h]
 \centering
 \includegraphics[width=1.00\textwidth]{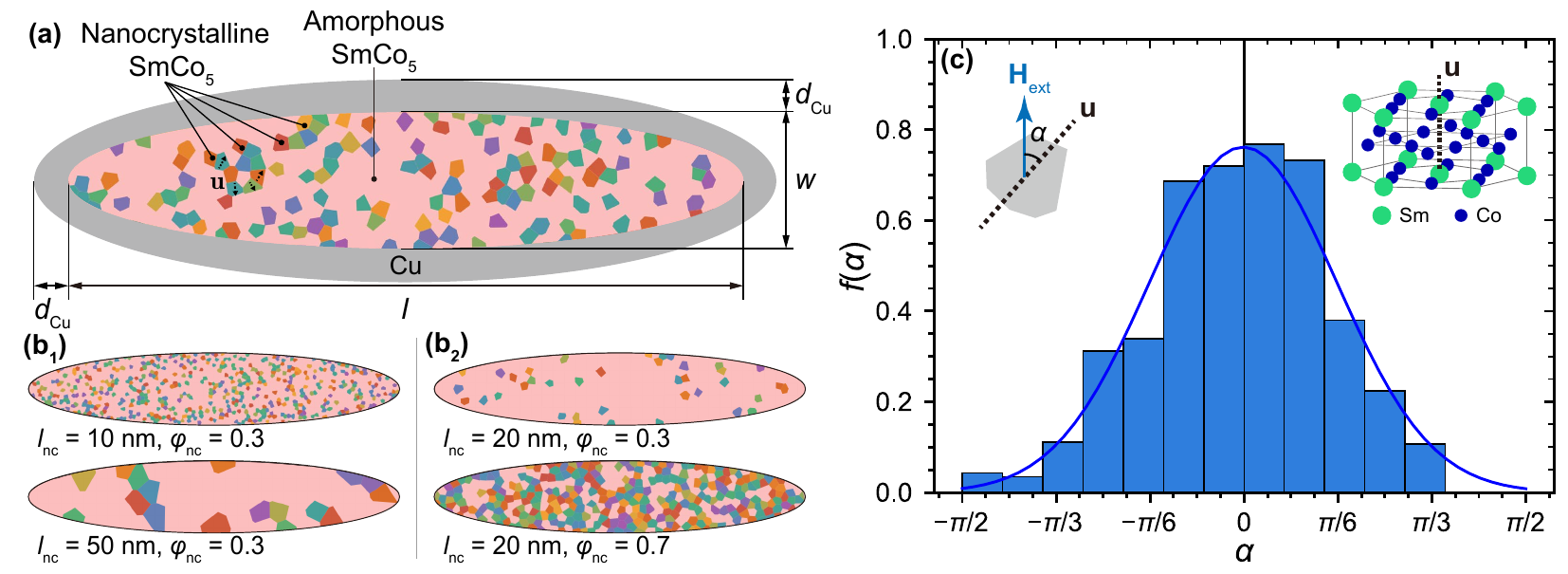}
 \caption{(a) Schematic of the parameterized elliptical nanocomposites for micromagnetic simulations, containing \ce{SmCo_5} nanocrystals and the amorphous matrix. Two varying parameters are focused: (b$_1$) varying average nanocrystal size $\ngl$ from 10 to 50 \si{nm} with fixing volume fraction $\fv=0.3$; (b$_2$) varying $\fv$ from 0.1 to 0.7 with $\ngl=20$ \si{nm}. (c) Pre-scribed orientation angle ($\alpha$) distribution for random-assignment of the nanocrystalline easy axis ($\eau$). The probability density function $f(\alpha)$ takes the Gaussian form. The histogram of assigned $\alpha$ in the nanocomposite $\ngl=20~\si{nm}$, $\fv=0.3$ is also illustrated as an instance. Inset: definition of the angle $\alpha$ and schematic of the \ce{SmCo5} unit cell with $\eau$ denoted \cite{gutfleisch2009high}. }
\label{Fig_Simulationen}
\end{figure*}

To understand the saturation of coercivity for the sample after 20 rotations (whereas the \ce{SmCo_5} phase is within the single domain particle size), the role of the amorphous phase and the embedded nanocrystals is studied by micromagnetic simulation based on experimental characterization. In this regard, we consider the simulation domains containing an elliptical nanocomposite region with the major axis $l=900~\si{nm}$ and minor axis $w=180~\si{nm}$ (aspect ratio of 5:1), and a Cu-coated region with the thickness $d_\mathrm{Cu}=50~\si{nm}$, as shown in \figref{Fig_Simulationen}a. This is based on the experimental characterization results shown in \figref{Fig_Microstructure_Experiment}. Nanocrystalline and amorphous \ce{SmCo_5} inside the nanocomposite region were created via the Voronoi tessellation on the randomly-labeled seeds. To control the size and volume fraction of generated nanocrystals, seeds with a uniform-controlled diameter were sampled based on the fast Poisson disk sampling \cite{bridson2007fast}. The amount of the seeds labeled as nanocrystals was further constrained according to the imposed seed fraction. When the seeding amount is sufficiently large, the average nanocrystal size $\ngl$ and nanocrystal volume fraction $\fv$ can be then approximated by the seed diameter and the seed fraction. To investigate the effects of the nanocrystal size and of the volume fraction separately and quantitatively, $\fv$ was fixed at 0.3 for varying $\ngl$ from 10 to 50 \si{nm} with an increment of 10 \si{nm}. For varying $\fv$ from 0.1 to 0.7 with an increment of 0.2, $\ngl$ was fixed at 20 \si{nm}. In \figref{Fig_Simulationen}b we demonstrate the generated nanocomposite with either fixed $l_\mathrm{nc}$ or fixed $\varphi_\mathrm{nc}$. For reference, a fully packed nanocomposite (i.e., $\fv=1.0$) with a nanocrystal size of $l_\mathrm{nc}=20~\si{nm}$ was also examined. The easy axes $\mathbf{u}$ of the nanocrystals were also assigned by random sampling from a Gaussian distribution of the misorientation angle $\alpha$ to the applied field, as shown in \figref{Fig_Simulationen}c. The mean and the standard deviation of $\alpha$ are $\mu_\alpha = 0$ and $\sigma_\alpha=\pi/6$, respectively, which can guarantee $99.73\%$ of the sampled $\alpha$ in-between $-\pi/2$ and $\pi/2$. 

The free energy of the system with volume $V$ was formulated as the functional of $\mag(\pos)$ following the scenario of micromagnetics, i.e.,
\begin{equation}
    \mathcal{F}=\int_V \left[f_\mathrm{ex}+f_\mathrm{ani}+f_\mathrm{ms}+f_\mathrm{zm}\right]\diff V.
    \label{eq:F}
\end{equation}
Here, $f_\mathrm{ex}$ is the exchange contribution, taking into account the parallel-aligning tendency between neighbouring magnetic moments due to the Heisenberg exchange interaction. $f_\mathrm{ani}$ represents the contribution of the magnetocrystalline anisotropy. The magnetostatic term $f_\mathrm{ms}$ counts the energy of each local magnetization under the demagnetizing field $\Hv_\mathrm{dm}$ created by the surrounding magnetization. The Zeeman term $\fed{zm}$ counts the energy of each local magnetization under an extrinsic magnetic field $\Hext$. Those terms are assigned to the distinctive phases, as shown in \tabref{raw}. The magnetic properties of nanocrystalline \ce{SmCo_{5}} employed in the simulation are $\Ae=8.6~\si{pJ/m}$, $\Ku=18.3~\si{MJ/m^3}$, and $\Msat=810.8~\si{kA/m}$ \cite{lectard1994saturation, coey2010magnetism}. 
Due to the lack of experimental investigations, the amorphous \ce{SmCo5} phase is assumed to have the identical exchange stiffness $\Ae$ and saturation magnetization $\Msat$ as the nanocrystalline phase, while its magnetocrystalline anisotropy is assumed to be uniaxial with $\Ku^\mathrm{am}=0.01\Ku=0.18~\si{kA/m}$. This is based on the experimental observations on similar hard magnetic systems, notably amorphous Nd-Fe-B \cite{zhang2013coercivity, harada1993production, kronmuller1981domains}, where the amorphous phase is treated as a soft magnetic matrix and is generally known to have negligible magnetocrystalline anisotropy \cite{kronmuller1981domains, kronmuller2003micromagnetism, fujimori1976soft}. 
The appearence of uniaxial magnetocrystalline anisotropy is due to local internal stresses or induced anisotropy by applied fields, which explains the non-zero $\Ku^\mathrm{am}$ for the amorphous phase. The exchange length is evaluated as $\lex = 2.15~\si{nm}$ in the nanocrystalline phase and $21.5~\si{nm}$ in the amorphous phase.
As for the diamagnetic Cu-phase, magnetic susceptibility $\chi=-6\E{-5}$ is utilized for physical consistency, though the Cu-phase has a negligible contribution to the magnetization reversal of the whole nanostructures.
As unveiled by microstructure analyses in \figref{Fig_Microstructure_Experiment}a, \ce{SmCo_{5}} particles are surrounded by the Cu-phase, implying a magnetic decoupling among particles.
%
\begin{table*}[]
\centering
\caption{Free energy density terms of corresponding phases. }
\begin{tabularx}{0.9\textwidth}{lcccc}
\hline
                            & $f_\mathrm{ex}$                          & $f_\mathrm{ani}$       & $f_\mathrm{ms}$                                      & $f_\mathrm{zm}$                            \\
\hline
\ce{SmCo_5} (nanocrystal) & \multirow{2}{*}{$\Ae\sum_i|\nabla m_i|^2$} & $-\Ku (\mag\cdot\eau)^2$ & \multirow{2}{*}{$-\frac{1}{2}\mu_0\Msat\mag\cdot\Hdm$} & \multirow{2}{*}{$-\mu_0\Msat\mag\cdot\Hext$} \\
\ce{SmCo_5} (amorphous)   &                                          & $-\Ku^\mathrm{am} (\mag\cdot\eau)^2$                       &                                                      &                                            \\ 
\ce{Cu}                   & 0                                        & 0                       & $-\frac{1}{2}\mu_0\chi\hext\mag\cdot\Hdm$                                                      & $-\mu_0\chi\hext\mag\cdot\Hext$   
\\ 
\hline
\end{tabularx}
\label{raw}
\end{table*}

The magnetization reversal under an imposed cycling magnetic field was generally described by the Landau-Lifshitz-Gilbert (LLG) equation. However, due to the incomparable time scale of LLG-described magnetization dynamics (around nanoseconds) with respect to the one of hysteresis measurement (around seconds), constrained optimization of the free energy functional $\mathcal{F}$ has been widely employed as a computationally efficient alternative to the time-dependent calculation in evaluating the hysteresis behavior of permanent magnets, where the magnetization configuration $\mv(\pos)$ is regarded as the quasi-equilibrium one \cite{Exl2014, schabes1988magnetization, furuya2015semi}. Based on the steepest conjugate gradient (SCG) method, it has been shown that the evolving direction of $\mv$ to optimize $\mathcal{F}$ is in accordance with the exact damping term of LLG equation \cite{Exl2014, furuya2015semi}, i.e., 
\begin{equation}
\begin{split}
&\frac{\partial\mv}{\partial t} = - \mv \times\left(\mv \times \dfedfdm \right) , \\
&\text{subject to}\quad|\mv|=1. \label{eq:gov_mm}
\end{split}
\end{equation}
The micromagnetic simulations were carried out by the FDM-based SCG minimizer in the open-sourced package MuMax$^3$ \cite{Vansteenkiste2014} with numerical details elaborated in Refs. \cite{Exl2014, berkov1993solving}.
The simulation domains were constructed with the $1024\times512\times2$ finite difference grids and a grid size of $1\times1\times10~\si{nm^3}$. Periodic boundary condition (PBC) was applied on the boundaries perpendicular to $z$-direction by macro geometry approach \cite{fangohr2009new}, while Neumann boundary condition was applied on other boundaries \cite{Vansteenkiste2014}. It should be notified that the simulation domain is equivalent to a long elliptic cylindrical structure with columnar nanocrystals due to the applied PBC. In order to take numerical fluctuations into account, five cycles of the hysteresis were examined for each nonstructure/reference, with the averaged one presented in the following contents.

\begin{figure*}[h]
 \centering
 \includegraphics[width=1.00\textwidth]{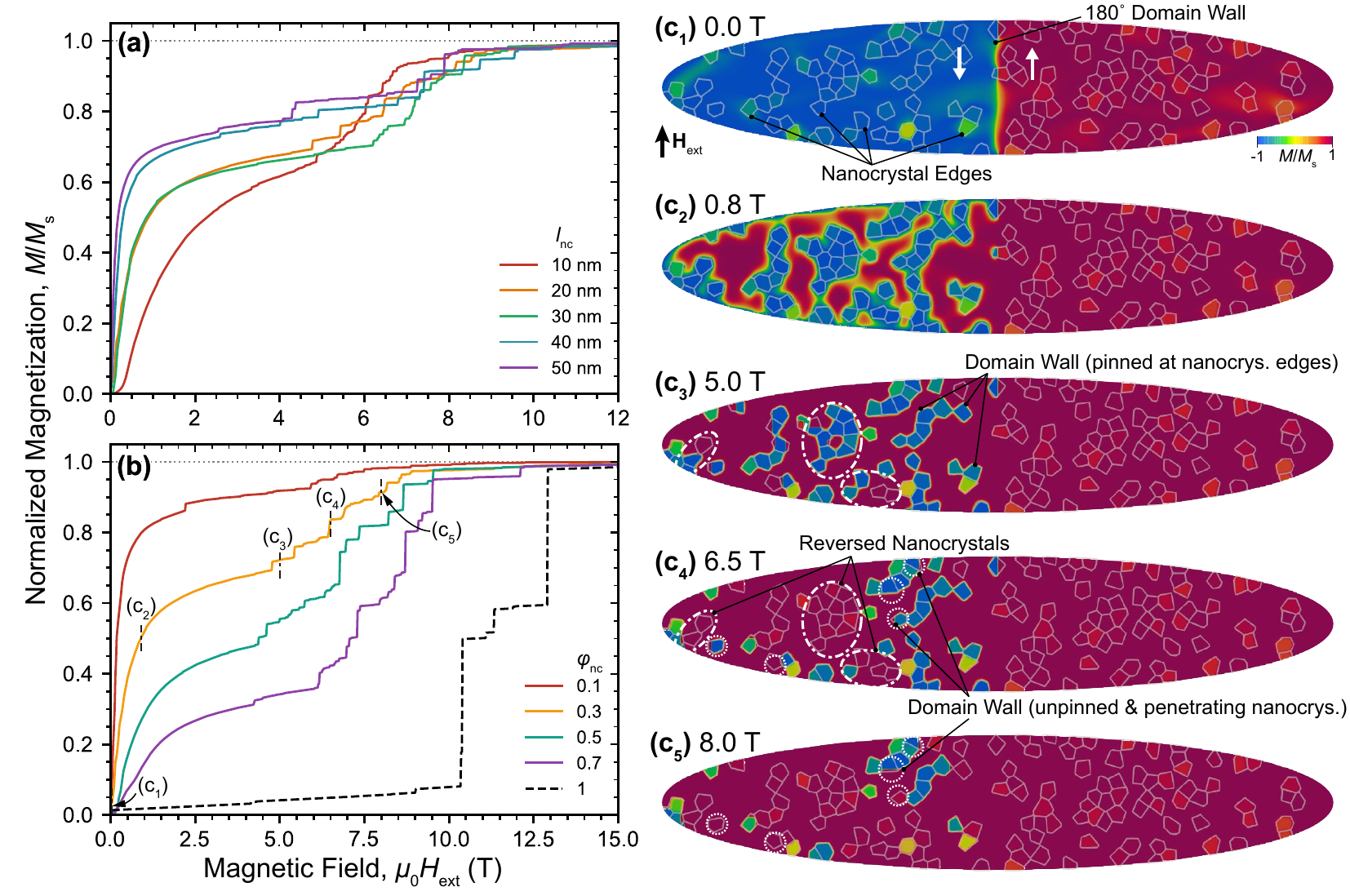}
 \caption{Simulated initial hysteresis curves for (a) varying nanocrystal size $\ngl$ with constant volume fraction $\fv=0.3$; (b) varying $\fv$ with constant $\ngl=20~\si{nm}$; (c) micromagnetic models for $\fv=0.3$ and $\ngl=20~\si{nm}$ at specific magnetic fields of $\hext$ equal to (c\textsubscript{1}) 0.0 \si{T}, (c\textsubscript{2}) 0.8 \si{T}, (c\textsubscript{3}) 5.0 \si{T}, (c\textsubscript{4}) 6.5 \si{T} and (c\textsubscript{5}) 8.0 \si{T}.}
\label{fig:init}
\end{figure*}

\figref{fig:init}a shows the initial magnetization curves for varying nanocrystal size with a constant volume fraction of $\fv=0.3$ whereas \figref{fig:init}b shows it for varying volume fraction with the constant size of the nanocrystals of $\ngl=20~\si{nm}$.
To get a precise understanding how the initial magnetization occurs in this microstructure exhibiting an amorphous matrix with embedded nanocrystals, when applying a field, the micromagnetic simulation of the magnetization processes at different external fields are plotted. The micromagnetic model for one specific case of $\fv=0.3$ and $\ngl=20~\si{nm}$ can be seen in \figref{fig:init}c. In the initial state half of the particle is magnetized upwards and half of the particle is magnetized downwards (\figref{fig:init}c).

\figref{fig:init}a and b show that the saturation for all different cases is achieved at $\hext=13~\si{T}$. From the microstructure (\figref{fig:init}c) it is evident that first the amorphous phase turns its magnetization direction towards the external field at very small applied fields before the magnetization of the nanocrystals reverses. The continuous magnetization of the initial curve in the beginning stems from the amorphous phase. At the boundaries of nanocrystals the domain wall gets pinned, which causes the jumps of the magnetization curve. With increasing external field, the domain wall gets unpinned at the boundaries of the nanocrystals and penetrates those. If the magnetization of a nanocrystal points in the direction of the external field the domain grows within a group of neighbouring crystallites. The jumps originate from single nanocrystals or groups of neighbouring nanocrystals which change their magnetization with increasing field. 

\figref{fig:init}a indicates a change of the coercivity mechanism with decreasing size of the nanocrystals from nucleation to pinning dominant for $\ngl=10~\si{nm}$. The same can be seen in \figref{fig:init}b for a constant size of the nanocrystal  $\ngl=20~\si{nm}$, when having a fully crystallite particle without an amorphous phase. For small volume fractions of nanocrystals the coercivity mechanism is nucleation controlled and changes to pinning for $\fv=1.0$. With increasing volume fraction of the crystallites it is also obvious that higher fields are necessary to reach magnetization saturation.

\begin{figure*}[h]
 \centering
 \includegraphics[width=1.00\textwidth]{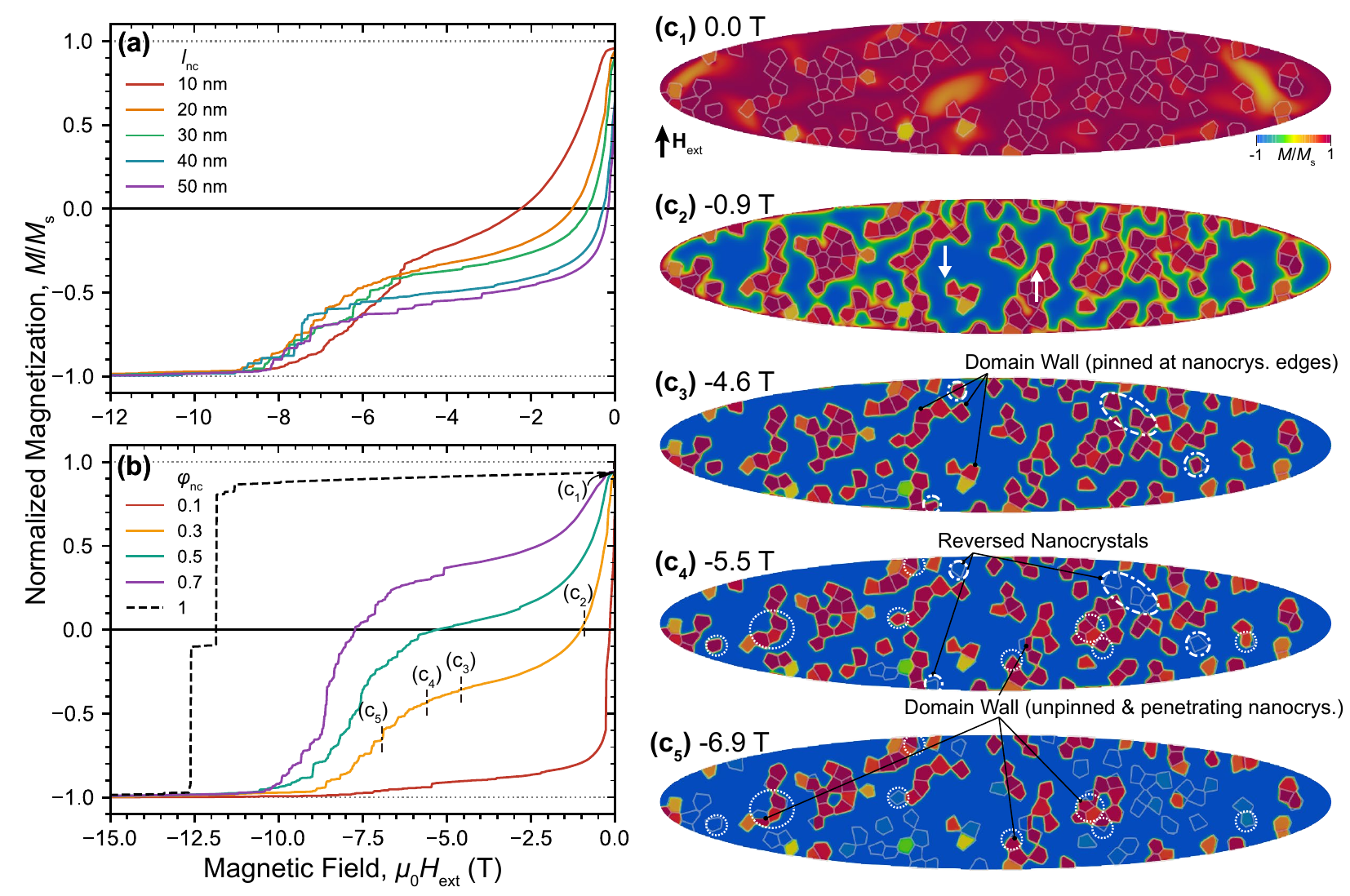}
 \caption{Simulated second and third quadrant of the hysteresis curves for (a) varying nanocrystallite size $\ngl$ with constant volume fraction $\fv=0.3$; (b) varying $\fv$ with constant $\ngl=20~\si{nm}$; (c) micromagnetic models for $\fv=0.3$ and $\ngl=20~\si{nm}$ at specific magnetic fields of $\hext$ equal to (c\textsubscript{1}) 0.0 \si{T}, (c\textsubscript{2}) -0.9 \si{T}, (c\textsubscript{3}) -4.6 \si{T}, (c\textsubscript{4}) -5.5 \si{T} and (c\textsubscript{5}) -6.9 \si{T}. Here $\Msat$ is valued as the maximum magnetization of each hysteresis.}
\label{fig:demag}
\end{figure*}

\begin{figure*}[h]
 \centering
 \includegraphics[width=1.00\columnwidth]{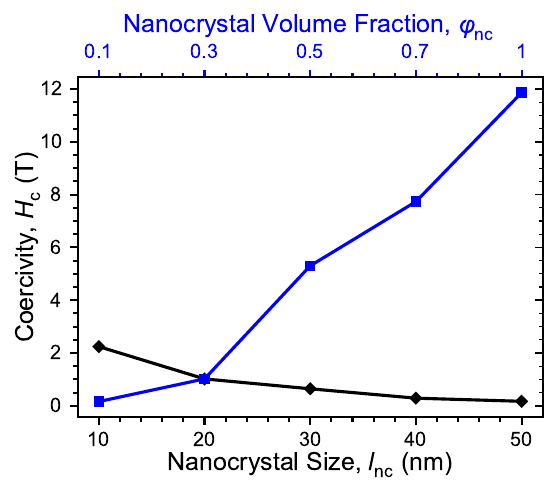}
 \caption{Simulated coercivity as a function of the nanocrystallite size $\ngl$ for a constant volume fraction of $\fv=0.3$, and as a function of the volume fraction $\fv$for a constant nanocrystallite size $\ngl=20~\si{nm}$. Here $\Msat$ is valued as the maximum magnetization of each hysteresis.}
\label{Fig_Hc}
\end{figure*}

When applying a negative field after saturation, the second and third quadrant of the hysteresis curve is depicted in \figref{fig:demag}a for varying nanocrystal size with a constant volume fraction of $\fv=0.3$. \figref{fig:demag}b shows it for varying volume fraction with constant size of the nanocrystals of $\ngl=20~\si{nm}$. The micromagnetic model can be seen again for different applied external field strengths for one specific case of $\fv=0.3$ and $\ngl=20~\si{nm}$ in \figref{fig:demag}c. \figref{fig:demag}a shows that a decreasing size of the nanocrystals leads to an increasing coercivity for a constant volume fraction of nanocrystals, since the number of crystal boundaries increases and hence pinning sites with decreasing size of nanocrystals.
\\ \\
By the aid of the micromagnetic model, it is obvious that when applying a negative field after magnetization of the particles (\figref{fig:demag}c\textsubscript{1}) first the magnetization of the amorphous phase rotates at very small negative fields starting at $-0.1 \si{T}$ (\figref{fig:demag}c\textsubscript{2} for $\hext=-0.9 \si{T}$), which corresponds to the smooth decrease of Magnetization (\figref{fig:demag}a). For the case of $\fv=0.3$ and $\ngl=20~\si{nm}$ the nanocrystals do not change their magnetization until a field of -5.2 \si{T} is reached, which is already much higher than its coercivity. Until this field, the domain wall is pinned at the nanocrystal boundaries. With further increasing negative field, first nanocrystals change their magnetization, domain walls are unpinned and penetrate into the nanocrystals and align them with the external field. The jumps originate again from single nanocrystals or groups of neighbouring nanocrystals which change their magnetization with increasing field in opposite direction. For a constant volume fraction of $\fv=0.3$ a decreasing size of the nanocrystals leads to higher coercivities, but for all cases magnetization reversal of the nanocrystals starts at negative fields which are larger than the coercivity, meaning that the pinning effects on the nanocrystals edges have no influence on the coercivity (\figref{fig:demag}a). For a constant size of the nanocrystals an increasing volume fraction of the nanocrystals leads to an increasing coercivity (\figref{fig:demag}b, \figref{Fig_Hc}). Especially for the case of $\fv=0.7$ the pinning effects on the nanocrystal edges lead to increased $\Hc$. The highest coercivity is reached if the particle does not contain amorphous phase $\fv=1.0$, which means that the amorphous phase is disadvantageous for the increasing coercivity. This can be explained by the magnetocrystalline anisotropy, which is significantly lower than for nanocrystals due to the lack of crystallinity of the amorphous phase.

In \figref{Fig_Hc} the coercivity is depicted as a function of the volume fraction of the nanocrystals $\fv$ (for constant $\ngl=20~\si{nm}$) and as a function of the nanocrystallite size $\ngl$ (for constant $\fv=0.3$). It is obvious that the influence of the volume fraction of nanocrystals is much more distinct compared to the influence of the nanocrystallite size. This is due to the fact that the pinning effects of the nanocrystal boundaries for the small volume fractions of nanocrystals only start at external field strengths which are higher than the coercivity.
\\ \\
Comparing the simulated coercivities with the experimental observations, they fit very well. Especially for the nanocrystalline volume fraction of $\fv=0.3$ and a size of the nanocrystals between $\ngl=10~\si{nm}$ and $\ngl=20~\si{nm}$ which is in good agreement with the microstructural observations for the small particles made by TEM (\figref{Fig_Microstructure_Experiment}d). By the aid of the micromagnetic simulations the effect of the amorphous phase and the embedded nanocrystals can be understood on a length scale at which an experimental investigation is very difficult. The simulations show that to obtain a high coercivity the amorphous phase is disadvantageous and domain wall pinning occurs at the boundaries of the nanocrystals. Except the particle containing large volume fractions of nanocrystals $\fv>0.7$, the pinning at nanocrystal boundaries has no effect on the coercivity since pinning occurs at negative external fields which are larger than the coercivity. Hence, the negative effect of the amorphous phase predominates. The formation of amorphous phase can also explain the strong decrease of the coercivity even though the microstructure gets refined further, when subjecting the sample to 100 rotations.
The results of the simulations, revealing the negative effect of the amorphous phase, show that the HPT process needs to be adjusted in such a way that the particle size of \ce{SmCo_{5}} is reduced to the single domain region while simultaneously the formation of amorphous phase is suppressed. One possible process parameter which can be used to adjust the microstructure is the process-temperature which may lead to dynamic crystallization during the process and can thus suppress the formation of amorphous phase. Our micromagnetic simulations provide an understanding of the mechanisms of the saturation behavior of magnetic hardening due to the induced strain by HPT of \ce{SmCo_{5}}-Cu nanocomposites on a length scale that is hard to analyse by experimental investigations and can be used for optimization strategies.

\section*{Acknowledgements}

Authors acknowledge the financial support of German Science Foundation (DFG) in the framework of the Collaborative Research Centre Transregio 270 (CRC-TRR 270, project number 405553726, sub-projects A01, A06, A08, Z01, Z02, Z-INF). Authors Y.Y., E.F. and B.-X.X. appreciate their access to the Lichtenberg High-Performance Computer and the technique supports from the HHLR, Technical University of Darmstadt, and the GPU Cluster from the sub-project Z-INF of SFB/TRR 270. Y. Y. also highly thanks Jiajun Sun, Zhejiang University, for his help in the technical check of the micromagnetic simulations.

\section*{Data Availability}
The authors declare that the data supporting the findings of this study are available within the paper. The microstructure generation scripts, micromagnetic input files, and utilities are cured in the online dataset (DOI: \url{xx.xxxx/zenodo.xxxxxxxx}).

\section*{Declaration of Competing Interest}

The authors declare that they have no known competing financial interests of personal relationships that could have appeared to influence the work reported in this paper.

\bibliography{reference}

\clearpage

\clearpage


\end{document}